\documentclass[11pt,twoside]{article}
\usepackage{hepro5}
\usepackage{graphicx}

\usepackage[T1]{fontenc} 

\usepackage{latexsym}
\usepackage{verbatim}

\begin{document}

\vskip 1.0cm
\markboth{C.~Pepe et al.}{Lepto-hadronic jet model for Cyg X-1}
\pagestyle{myheadings}

\vspace*{0.5cm}
\title{ Spectral energy distribution, polarization, and synthetic radio maps of Cygnus X-1: a lepto-hadronic jet model}

\author{C.~Pepe,$^1$ G.S.~Vila$^1$ and G.E.~Romero$^{1,2}$}
\affil{$^1$Instituto Argentino de Radioastronom\'ia (IAR-CONICET), C.C. 5, (1894) Villa Elisa, Buenos Aires, Argentina \\
$^2$Facultad de Ciencias Astron\'omicas y Geof\'isicas, Universidad Nacional de La Plata (FCAG-UNLP)\label{inst2}, Paseo del Bosque S/N, (1900) La Plata, Buenos Aires, Argentina}

\begin{abstract}
In this work we combine SEDs, radio maps and polarization observations
to understand the emission mechanisms in Cygnus X-1. Our radiative model
indicates that the MeV emission originates in the jet and that all the very
high-energy emission is from hadronic origin. We also performed a synthetic
radio map that suggests that our description of the magnetic field should be
improved, since it leads to a very compact emission region. In order to 
choose the most suitable magnetic field geometry, we investigated the polarization
in X-rays and we found that very simple geometries can explain the high levels
of polarization reported by other authors. 

\end{abstract}

\section{Introduction}

Cygnus X-1 (Cyg X-1) is probably the most widely monitored microquasar (MQ) in the Galaxy. The binary system, located at $1.86~\rm{kpc}$ from Earth (Reid et al. 2011), is composed of a high-mass star (spectral type O9.7 Iab and mass 
\mbox{$\sim 20~M_{\odot}$}) and a black hole of \mbox{$14.8~M_{\odot}$} (Orosz et al. 2011). 

A very complete broadband spectral energy distribution (SED) is available for Cyg X-1 in the hard state (for a compilation of the data see Zdziarski et al. 2014), including gamma-ray detections
and upper limits at GeV energies and above (Albert et al. 2007, Malyshev et al. 2013). The origin of the soft gamma rays ($\sim$ MeV), in particular, is still unknown: 
there is no agreement about whether they originate in the jets or somewhere else in the accretion flow. This is one of the issues we assess in this work.

Jets in Cyg X-1  have been resolved in the radio band (Stirling et al. 2001). The outflow is extremely collimated and mildly relativistic. The extension and geometry of the radio emission region may provide complementary, useful information about the conditions in the jets, such as the size and location of the acceleration region of relativistic particles, and the magnetic field. 

Finally, while polarization data at low energies have been long available for Cyg X-1 (see Russell et al. 2013 for a compilation),  high levels of polarization in the X rays/soft gamma rays have been measured recently for the first time (Laurent et al. 2011, Rodr\'iguez et al. 2015). Polarization studies of the jet radiation can help settle the issue of the origin of the MeV tail.

In this work, we combine these three different sources of data (non-thermal SED, radio images and polarization measurements) to obtain information about the conditions in the jets of Cyg X-1. In Sections \ref{radiative} and \ref{maps}, we briefly review the radiative model - developed in detail in Pepe et al. (2015) - and its application to the generation of synthetic radio maps. In Section \ref{polarization}, we
present our preliminary results for the degree of polarization of the jet synchrotron emission. Finally, in Section \ref{Conclusions} we discuss our conclusions and perspectives for future work.

\section{Radiative processes}
\label{radiative} 

\begin{figure}
\begin{center}
\hspace{0.25cm}
 \includegraphics[width = 0.5\textwidth, keepaspectratio]{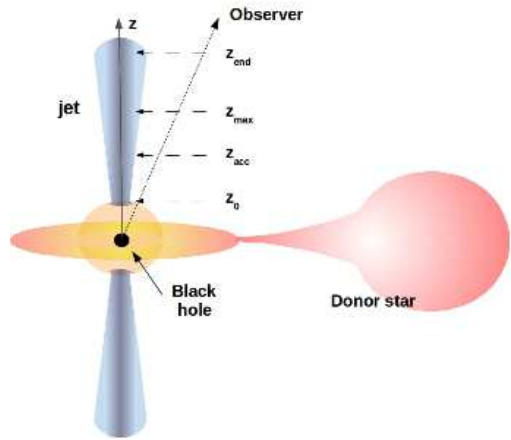}
 \caption{Basic sketch of the binary and the jet (not to scale). }
 \label{fig:sketch}
 \end{center}
\end{figure}

In this section we describe the modelling of the radiative output of Cyg X-1. The reader is referred to Pepe et al. (2015) for details. We adopt a conical
shape for the jet, see Fig. \ref{fig:sketch}. The jet base is located at a distance $z_0 = 1.1\times10^8$~cm from the compact object. Relativistic particles are injected in a region that
starts at $z_{\rm{acc}}= 2.2\times10^8$~cm and extends up to $z_{\rm{max}}= 8.6\times10^{11}$~cm. The jet ends (for computing purposes) at $z_{\rm{end}}= 1.0\times10^{15}$~cm. The magnetic field at the base, $B_0 = 5.0\times10^7$~G, is estimated from equipartition between the magnetic and kinetic energies and it decays as $B(z) = B_0 (z_0/z)$. Given the total power of the jet $L_{\rm{jet}}$, a power

\begin{equation}
 L_{\rm{rel}} = q_{\rm{rel}} L_{\rm{jet}} \qquad q_{\rm{rel}} = 0.1
\end{equation}

\noindent is transferred to the relativistic particles in the acceleration region, which, in turn, is distributed between electrons and protons as

\begin{equation}
 L_{\rm{p}} = a L_{\rm{e}} \qquad a = 0.07.
\end{equation}

Relativistic particles are injected in the jet according to the injection function 

\begin{equation}
 Q(E,z) = Q_0\,  E^{-\Gamma}\, \exp{\left[-E/E_{\rm{max}}(z)\right]}.
\end{equation}

\noindent Here, $Q_0$ is a normalization constant obtained from the total power injected in each particle species and $\Gamma$ is the spectral index. The injection function is different from zero only in
the region $z_{\rm acc}\leq z \leq z_{\rm max}$ and for $E\geq E_{\rm{min}}$. The cutoff energy $E_{\rm{max}}$ is calculated equating the total particle energy loss rate and the acceleration
rate (e.g. Aharonian 2004); we adopt $E_{\rm{min}} = 120 m_0$, where $m_0$ is the rest mass of the particles.

Radiative cooling is calculated for all particles. Leptons cool via synchrotron, relativistic Bremsstrahlung and 
inverse Compton. For this last process we consider three different photon targets: electron synchrotron radiation (SSC), the radiation field of the companion star (IC-Star) and the X-ray photons from the accretion disk (IC-Disk). Protons cool via synchrotron, proton-proton ($pp$) and proton-photon ($p\gamma$) interactions.  In the 
case of $pp$ collisions the targets are the thermal protons in the jet and in the stellar wind ($pp$-Star), while the photons for $p\gamma$ interactions are those of the radiation field of the companion star. In the case of electrons, synchrotron losses are dominant until almost the end of the jet, while in the case of protons adiabatic losses govern the cooling nearly all along the jet. 

Once the cooling rates and the injection functions are calculated, we solve the steady-state transport equation for the particle distributions $N(E,z)$,

\begin{equation}
 v_{\rm{conv}} \frac{\partial N}{\partial z} + \frac{\partial}{\partial E} \left( \left.\frac{dE}{dt}\right|_{\rm{tot}} N \right) = Q(E,z),
 \label{eq:transport}
\end{equation}

\noindent for both protons and electrons. The main feature of this equation is that it accounts for the transport of particles with a convection velocity on the order of the jet bulk velocity, $v_{\rm{conv}} \approx v_{\rm{jet}}$. 

\begin{figure}[htbp]
 \begin{center}
  \includegraphics[width = 0.48\textwidth, keepaspectratio]{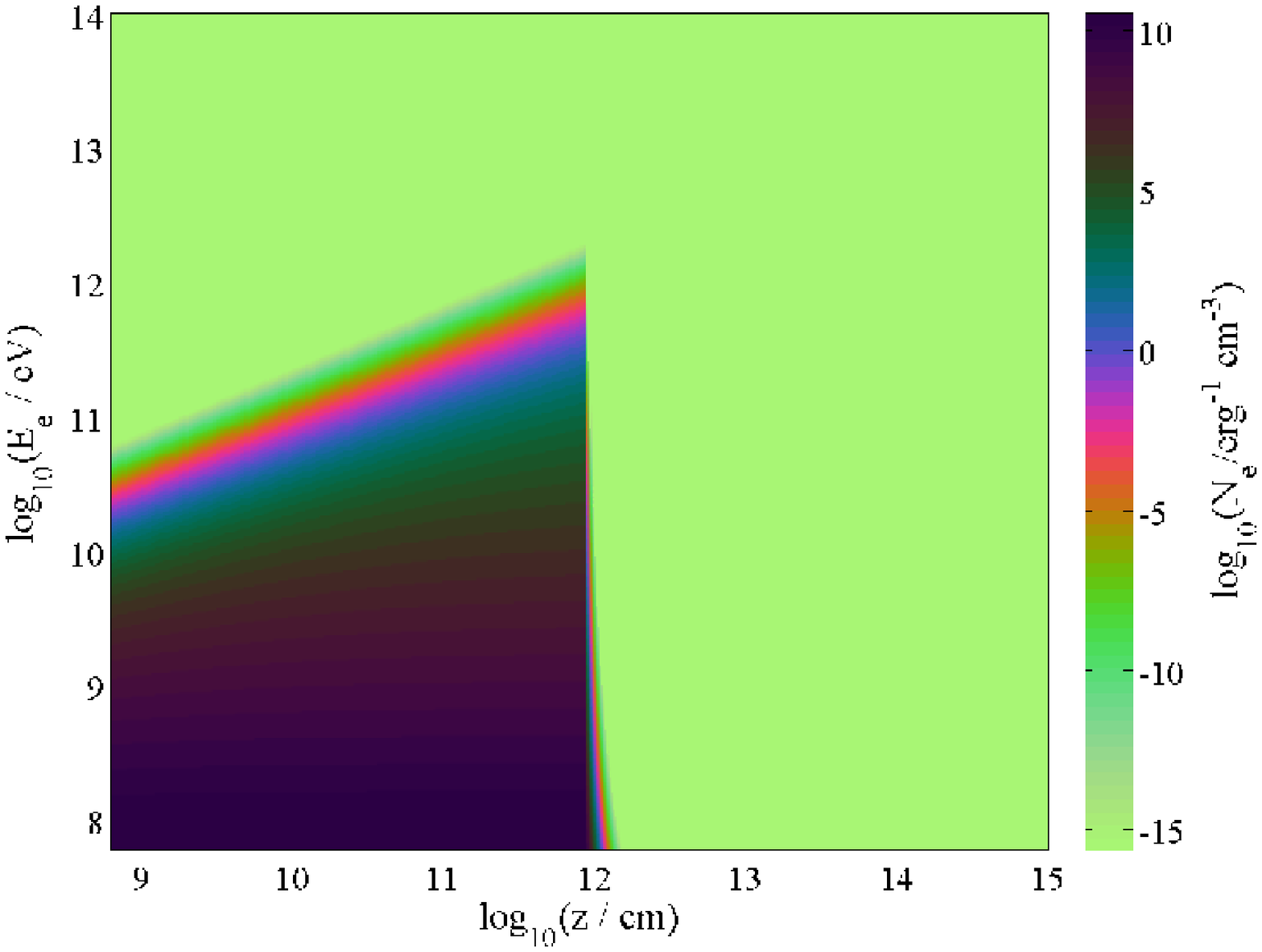}
  \includegraphics[width = 0.48\textwidth, keepaspectratio]{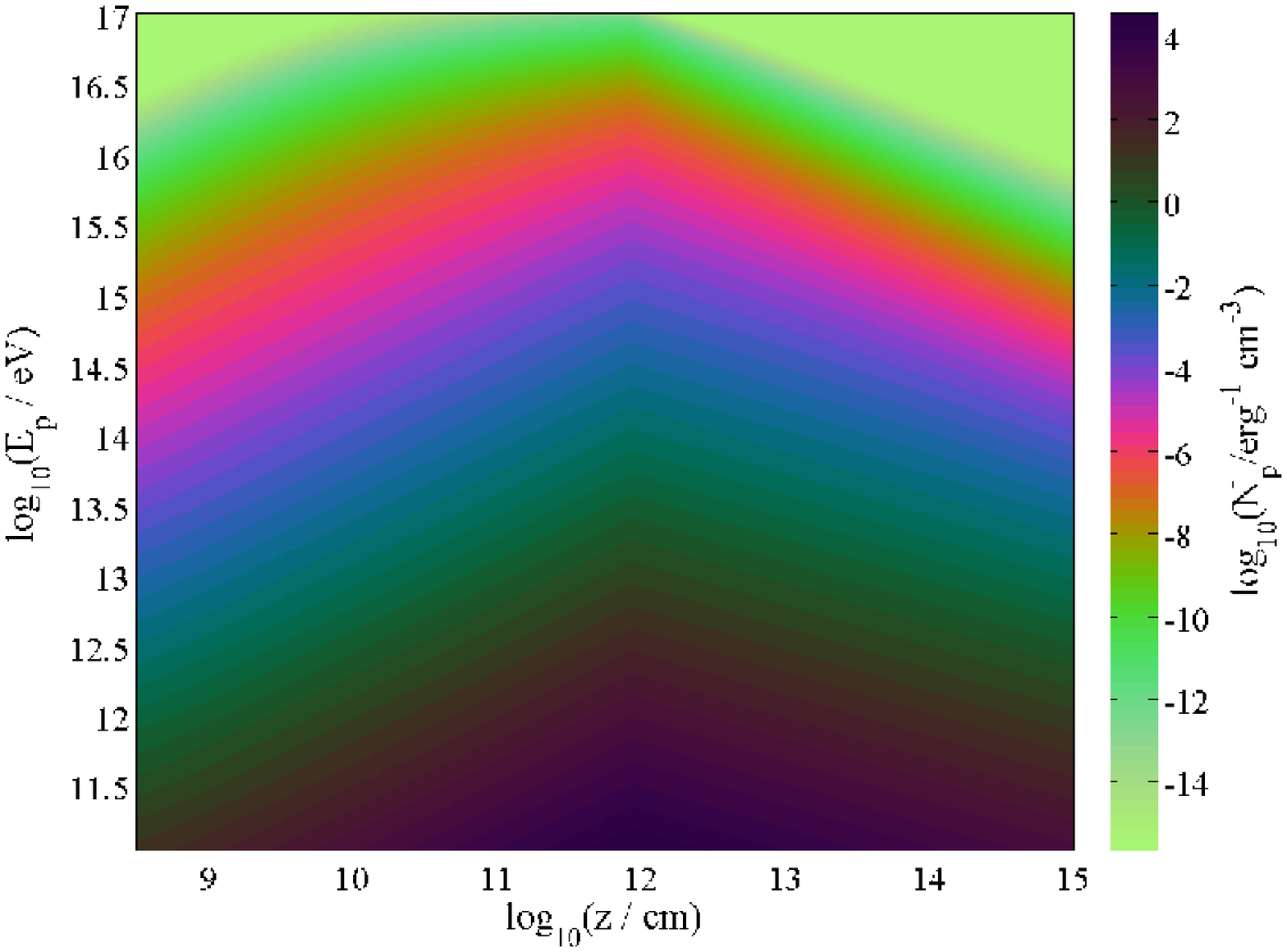}
 \caption{Steady-state distribution of electrons (left) and protons (right). }
 \label{fig:distribution}
  \end{center}
\end{figure}

The resulting electron and proton distributions are shown in Fig. \ref{fig:distribution}. Energetic protons can 
be found well outside the acceleration region, but electrons cool almost immediately after they leave the acceleration region.

\begin{figure*}[htp]
 \centering
 \includegraphics[width = 0.95\textwidth, keepaspectratio]{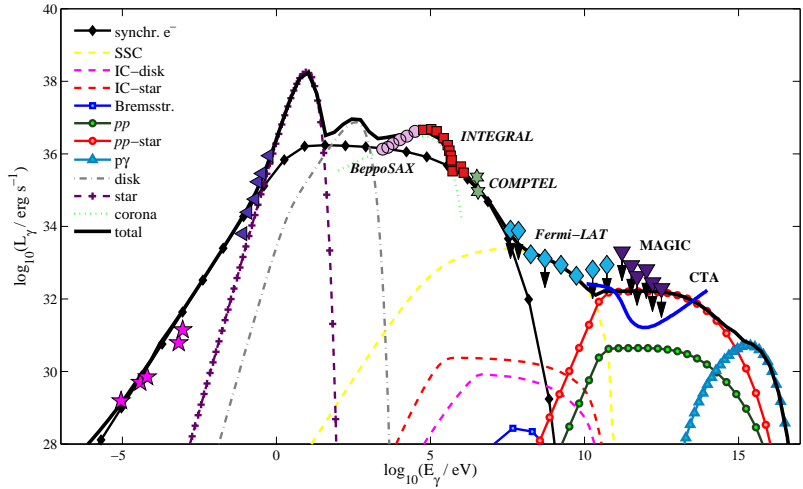}
 \caption{Best-fit spectral energy distribution for Cygnus X-1.  Down-pointing arrows indicate upper limits. The data are not simultaneous. See Pepe et al. (2015) for details on the sources of the data.}
 \label{fig:SED}
\end{figure*}

Once the particle distributions are known, we then obtain the 
specific luminosity in the observer reference frame (in $\rm{erg~s^{-1}~sr^{-1}}$) at photon energy $E_\gamma$ as

\begin{equation}
L_\gamma (E_\gamma) = E_\gamma\, \int_{V} q_\gamma\, dV  ,
\end{equation}

\noindent where $V$ is the volume of the emission region and  $q_\gamma$ the volume
emissivity. In Fig. \ref{fig:SED} we show our best-fit SED as well as the broadband data for Cyg X-1. In this model, the non-thermal emission from radio wavelengths to the the MeV tail is well described as synchrotron radiation from electrons in the jet. Note that all the emission above 10~GeV is exclusively
of hadronic origin. Furthermore, it is very close to the detections limits of MAGIC and CTA. If this emission were detected, it would be an indicator of the presence of protons in the outflows of Cyg X-1.\footnote{So far, heavy nuclei have been detected in the jets of only two MQs: SS~433 and 4U~1630C47.} 


\section{Radio maps}
\label{maps}

\begin{figure*}
\centering
 \includegraphics[width = 0.48 \textwidth, keepaspectratio]{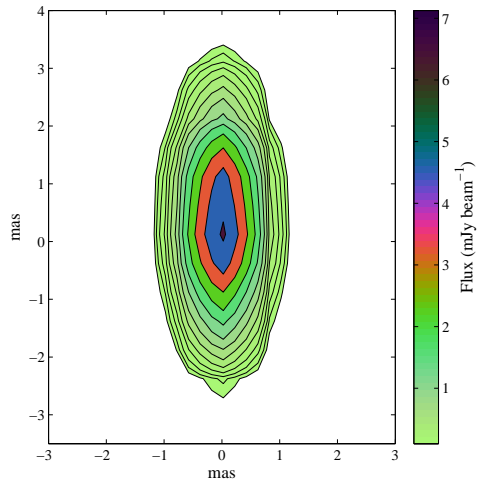}
 \caption{Image of the jet at $8.4$ GHz after convolution with a Gaussian beam of $2.25\times 0.86$ mas$^2$. The origin of coordinates is chosen to coincide with the position of the flux maximum. Contours are spaced in factors of $\sqrt{2}$; the lowest contour corresponds to 0.1~mJy~beam$^{-1}$.}
 \label{fig:mapas}
\end{figure*}

In this section we describe the procedure and results of our modelling of the radio emission region. We integrate the volume emissivity $q_\gamma$ 
along the line of sight and then convolve it with a bidimensional Gaussian function of full width at half maximum (FWHM) of $2.25\times 0.86$ mas$^2$ to mimic the effect of an array 
with a beam as in Fig. 3 of Stirling et al. (2001); the chosen separation between pointings was of one beam radius in each direction. The result is shown in Fig. \ref{fig:mapas}. The flux levels are comparable to those measured by Stirling et al. (2001);  the extension of the emitting region, however, is smaller. 
This may be an indication that our modelling of the acceleration region (size and/or position) and/or the magnetic field should be revised. 


\section{Synchrotron polarization}
\label{polarization}

\begin{figure}[htbp]
 \begin{center}
  \includegraphics[width = 0.18\textwidth, trim=0 0 0 0, clip]{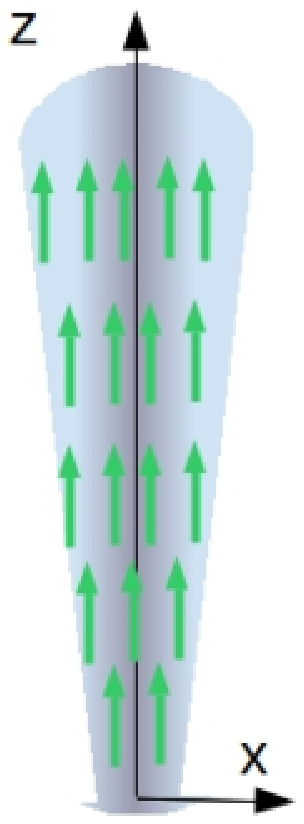}
  \includegraphics[width = 0.18\textwidth, trim=0 0 0 0, clip]{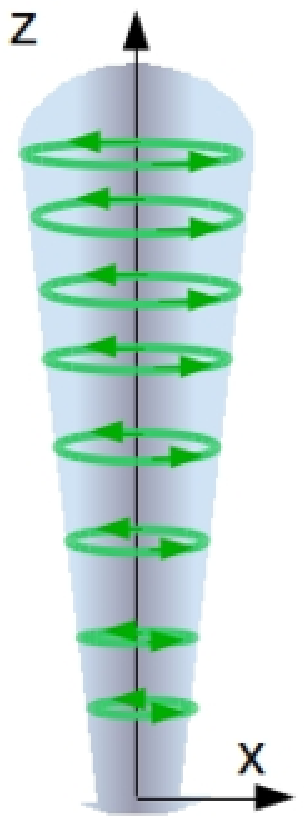}
 \caption{Magnetic field geometries explored: decaying field in the vertical direction (left, $B_z$) and toroidal magnetic field (right, $B_{\phi}$). The $z$ direction is parallel to the jet axis. }
 \label{fig:magneticField}
  \end{center}
\end{figure}

Polarization depends directly on the magnetic field strength and configuration. Hence, we study the polarization of the emitted radiation as a means of testing our description of the magnetic field in the jet. We particularly focus on the polarization of the MeV radiation in order to compare our results with the recent measurements in that energy range. We follow Korchakov \& Sirovatskii (1962) for the calculation of the degree of polarization. We 
compute the Stokes parameters from first principles, i.e., for completely general shapes of the magnetic field and particle distributions. We explore two different, simple geometries for the magnetic field; see Fig. \ref{fig:magneticField}. In both scenarios the magnetic field intensity decays with the coordinate $z$ as stated in Section \ref{radiative}. Our calculations indicate levels of polarization $\rho_{{B_{z}}} \sim 80\%$ and 
$\rho_{{B_{\phi}}} \sim 75\%$, comparable to those reported by Laurent et al. (2011) and Rodriguez et al. (2013).

\section{Conclusions}
\label{Conclusions}

In this paper we present our latest results for the modelling of the broadband emission of Cyg X-1. Our model indicates that the most energetic emission of this source is dominated by hadronic processes and that the MeV tail has a leptonic origin in the jets. We also obtain a flux of synchrotron radio emission consistent with 
observations. However, the compactness of the synthetic radio source indicates that our description of the magnetic field and/or the acceleration zones needs to be improved. Our first investigations of the polarization of the radiation in the MeV band show that very simple field geometries can account for the observed level of polarization. In future works, we expect to further exploit the radio and polarization measurements to improve our modelling of the jets. In this regard, we will improve our description of the jet magnetic field by considering more realistic geometries and explore other configurations for the particle (re-)acceleration zones.

\end{document}